\documentstyle[11pt,mrs2001,epsfig]{article}
\def \Mpc {~h^{-1}~{\rm Mpc} }
\def \Om {\Omega_0}
\def \Omm {\Omega_{\rm m}}

\def \lo {\lambda_0}
\def \kms {{\rm ~km~s}^{-1}}
\def \bj {b_{\rm J}}
\def \qso {_{\rm Q}}
\def \invMpc {h^{-1}~{\rm Mpc}}
\def \invMpcc {h^{-3}~{\rm Mpc}^3}
\def \sigp {\sigma_{\rm p}}

\begin{document}
\renewcommand{\floatpagefraction}{0.70}
\title{Clustering in the 2dF QSO Redshift Survey}

\author{S.M. Croom$^1$, B.J. Boyle$^1$, N.S. Loaring$^2$,
      L. Miller$^2$, P. Outram$^3$, T. Shanks$^3$, R.J. Smith$^4$,
      F. Hoyle$^5$}   
\affil{$^1$Anglo-Australian Observatory, PO Box 296, Epping, NSW 2121,
      Australia}
\affil{$^2$Department of Physics, Oxford University, Keble Road, Oxford, OX1
      3RH, UK.}  
\affil{$^3$Physics Department, University of Durham, South Road, Durham,
      DH1 3LE, UK.}
\affil{$^4$Liverpool John Moores University, Twelve Quays House,
      Egerton Wharf, Birkenhead, CH41 1LD, UK}
\affil{$^5$Department of Physics, Drexel University, 3141 Chestnut
      Street, Philadelphia, PA 19104, USA}

\begin{abstract}
We present clustering results from the 2dF QSO Redshift Survey (2QZ)
which currently contains over 20,000 QSOs at $z<3$.  The two-point
correlation function of QSOs averaged over the entire survey
($\bar{z}\simeq1.5$) is found to be similar to that of local galaxies.
When sub-dividing the sample as a function of redshift, we find that
for an Einstein-de Sitter universe QSO clustering is constant (in
comoving coordinates) over the entire redshift range probed by the
2QZ, while in a universe with $\Om=0.3$ and $\lo=0.7$ there is a
marginal increase in clustering with redshift.  Sub-dividing the 2QZ
on the basis of apparent magnitude we find only a slight difference
between the clustering of QSOs of different apparent brightness, with
the brightest QSOs having marginally stronger clustering.  We have
made a first measurement of the redshift space distortion of QSO
clustering, with the goal of determining the value of cosmological
parameters (in partcular $\lo$) from geometric distortions.  The
current data do not allow us to discriminate between models, however,
in combination with constraints from the evolution of mass clustering
we find $\Omega_{\rm m}=1-\lo=0.23^{+0.44}_{-0.13}$ and
$\beta(z\sim1.4)=0.39^{+0.18}_{-0.17}$. The full 2QZ data set will
provide further cosmological constraints.
\end{abstract}

\section{Introduction}

The 2dF QSO Redshift Survey (2QZ)  aims to compile a homogeneous
catalogue of $\sim25000$ QSOs using the Anglo-Australian Telescope
(AAT) 2-degree Field facility (2dF) \cite{2dfpaper}.
This catalogue will constitute a factor of $\ga50$ increase in
numbers to a equivalent flux limit over previous data sets \cite{bsfp90}.
The main science goal of the 2QZ is to use QSOs as a probe of
large-scale structure in the  Universe over a range of scales
from 1 to $1000\Mpc$, out to $z\simeq3$.
These measurements have the potential to help determine the
fundamental cosmological parameters governing the Universe.  The
2QZ will also significantly advance our understanding of the AGN/QSO
phenomenon, by allowing us to carry out statistical studies of large
numbers of QSO spectra, and also discovering rare and extreme types of
AGN, for example broad-absorption-line QSOs or post-starburst QSOs
\cite{brother99}.  As well as QSOs, there are a large number of other
interesting sources being discovered including compact
narrow-emission-line galaxies, white dwarfs and cataclysmic
variables. 

In this paper we will concentrate on the statistical measurements of
QSO clustering.  The basic properties of the 2QZ are outlined in 
Section \ref{sec:survey}.  We then discuss various measurements
of large-scale structure including the correlation function (Section
\ref{sec:xir}) and redshift-space distortions (Section
\ref{sec:zspace}).

\section{The 2dF QSO Redshift Survey}\label{sec:survey}

The 2QZ currently (September 2001) contains 20573 QSOs below a
redshift of $z\simeq3$.  Observations will be completed in January
2002, by which time close to 25000 QSOs will have been observed.  A
large fraction of the data is already publicly available; the 2QZ ``10k
catalogue'' \cite{2qzpaper5} contains spectra of 11000 QSOs, and
almost 10000 other sources.  The spectra and catalogue are
available electronically at {\tt http://www.2dfquasar.org}.  The
analysis below mostly concerns this 10k catalogue.

QSO candidates were selected as being point sources and bluer than
the stellar locus, based on broad band $u\bj r$ magnitudes from
Automatic Plate Measuring (APM) facility measurements of UK Schmidt
Telescope (UKST) photographic plates.  The magnitude limits are
$18.25<\bj\leq20.85$.  The survey comprises 30 UKST fields, arranged
in two $75^{\circ}\times5^{\circ}$ declination strips centred in the
South Galactic Cap (SGC) at $\delta=-30^{\circ}$ and the North
Galactic Cap (NGC) at $\delta=0^{\circ}$ with RA ranges
$\alpha=21^h40$ to $3^h15$ and $\alpha=9^h50$ to $14^h50$
respectively.  The completed survey will cover approximately 740
deg$^2$ \cite{croom,smith,2qzpaper3}. 

Spectroscopic observations have been carried out using the 2dF
instrument at the AAT in conjunction with the 2dF Galaxy Redshift
Survey (2dFGRS) \cite{2dfgrs}, as the 2QZ and 2dFGRS areas cover the
same regions of sky.  Spectroscopic data are reduced using the 2dF
pipeline reduction system \cite{2dfman}.  The identification of QSO
spectra and redshift estimation was carried out using the {\small
AUTOZ} code written specifically for this project.  This program
compares template spectra of QSOs, stars and galaxies to the observed
spectra.  Identifications are then confirmed by eye for all spectra.
The mean spectroscopic completeness for the sample is 89 per cent.

\begin{figure}
\plotone{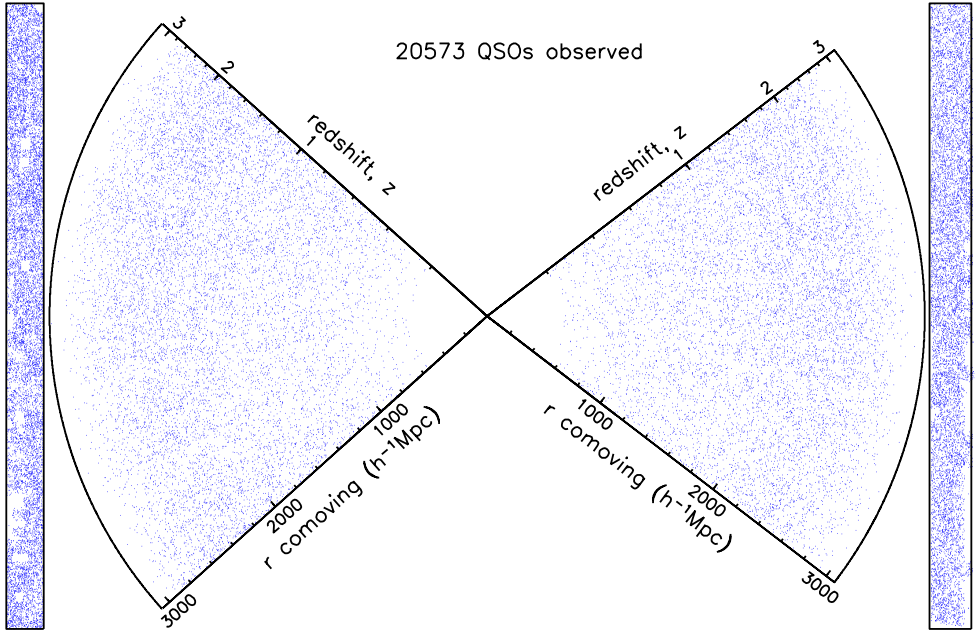}{14cm}
\caption{The spatial distribution of 2QZ QSOs as of September 2001.
the SGC strip is on the left, the NGC strip on the right.  The
rectangles at either end show the distribution of QSOs on the sky.  An
Einstein-de Sitter cosmology is assumed.}
\label{fig:wedge}
\end{figure}

The spatial and redshift distribution of QSOs is shown in Fig
\ref{fig:wedge}.  At redshifts $z\ga2$ the 
decline in QSO numbers is due to the increased reddening of QSO
colours by absorption in the Ly$\alpha$ forest.  QSOs in this
region are therefore missed by our colour selection.  At low redshifts
we will miss objects in which the host galaxy contributes
significantly to the flux, due to both the extended nature of the
sources and their redder colours.  An estimate of the
survey incompleteness due to the colour selection is given by Boyle et
al. (2000) \cite{2qzpaper1}.

\section{The QSO correlation function}\label{sec:xir}

\begin{figure}
\vspace{-0.9cm}
\plottwo{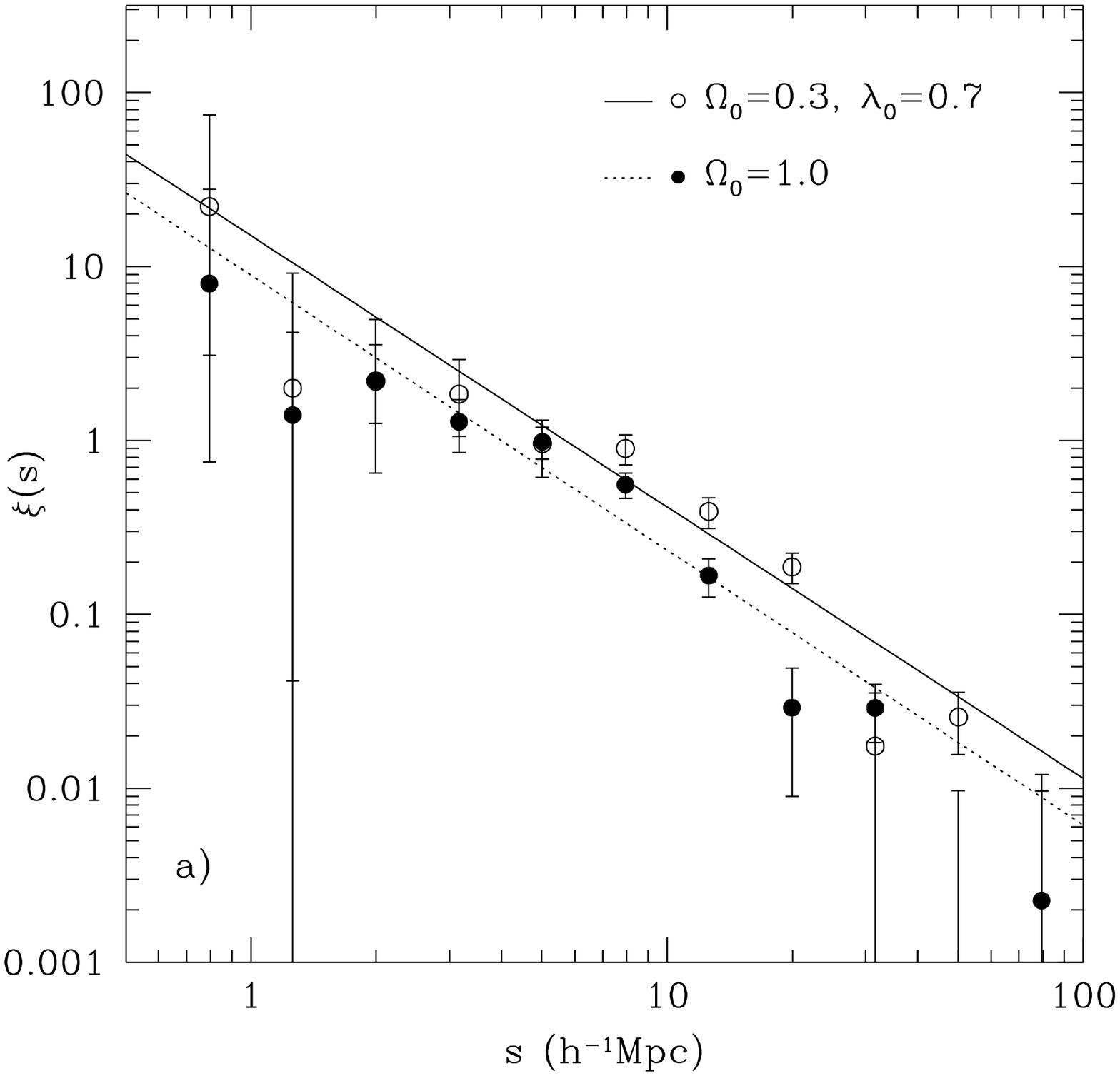}{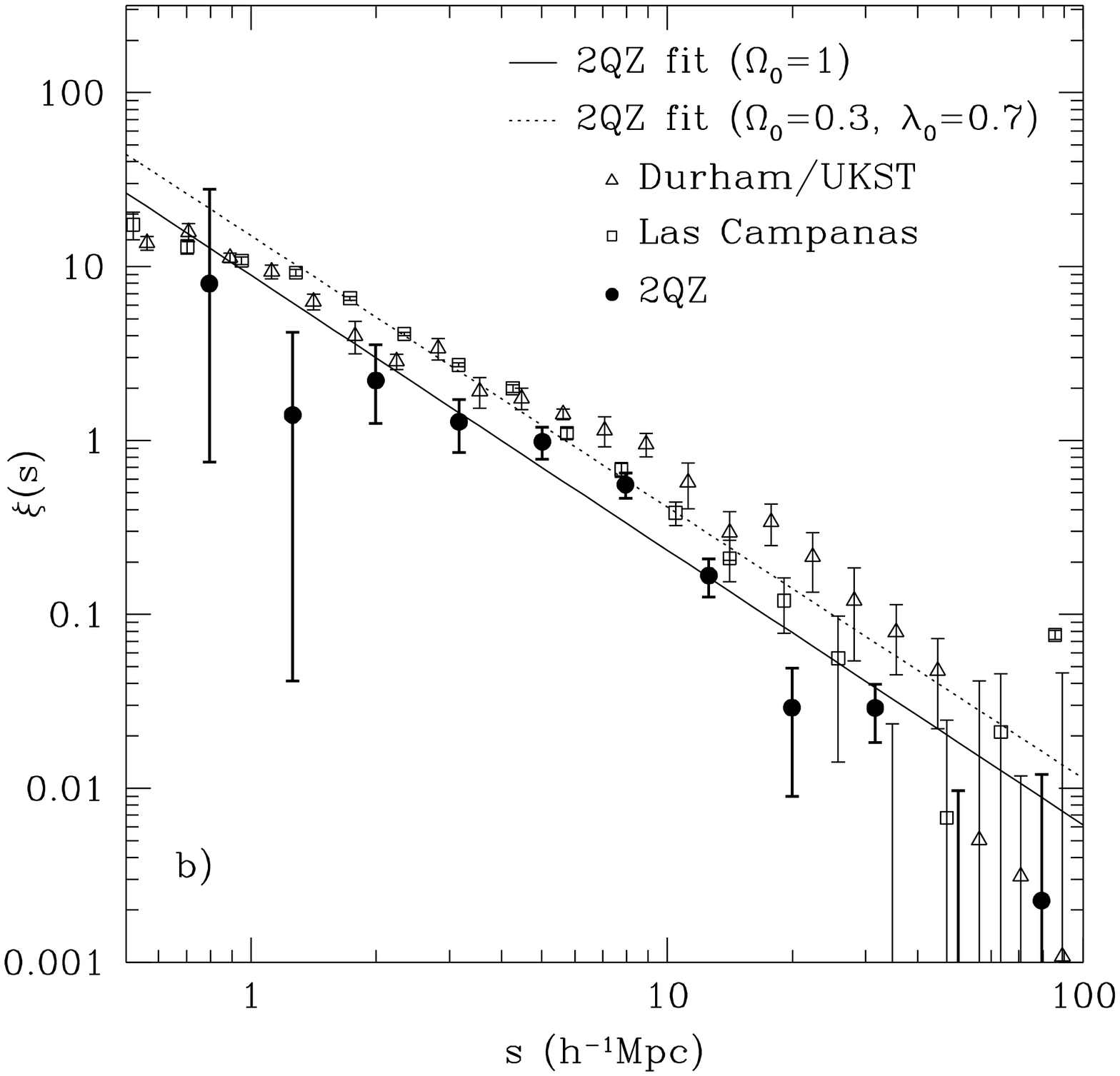}
\plottwo{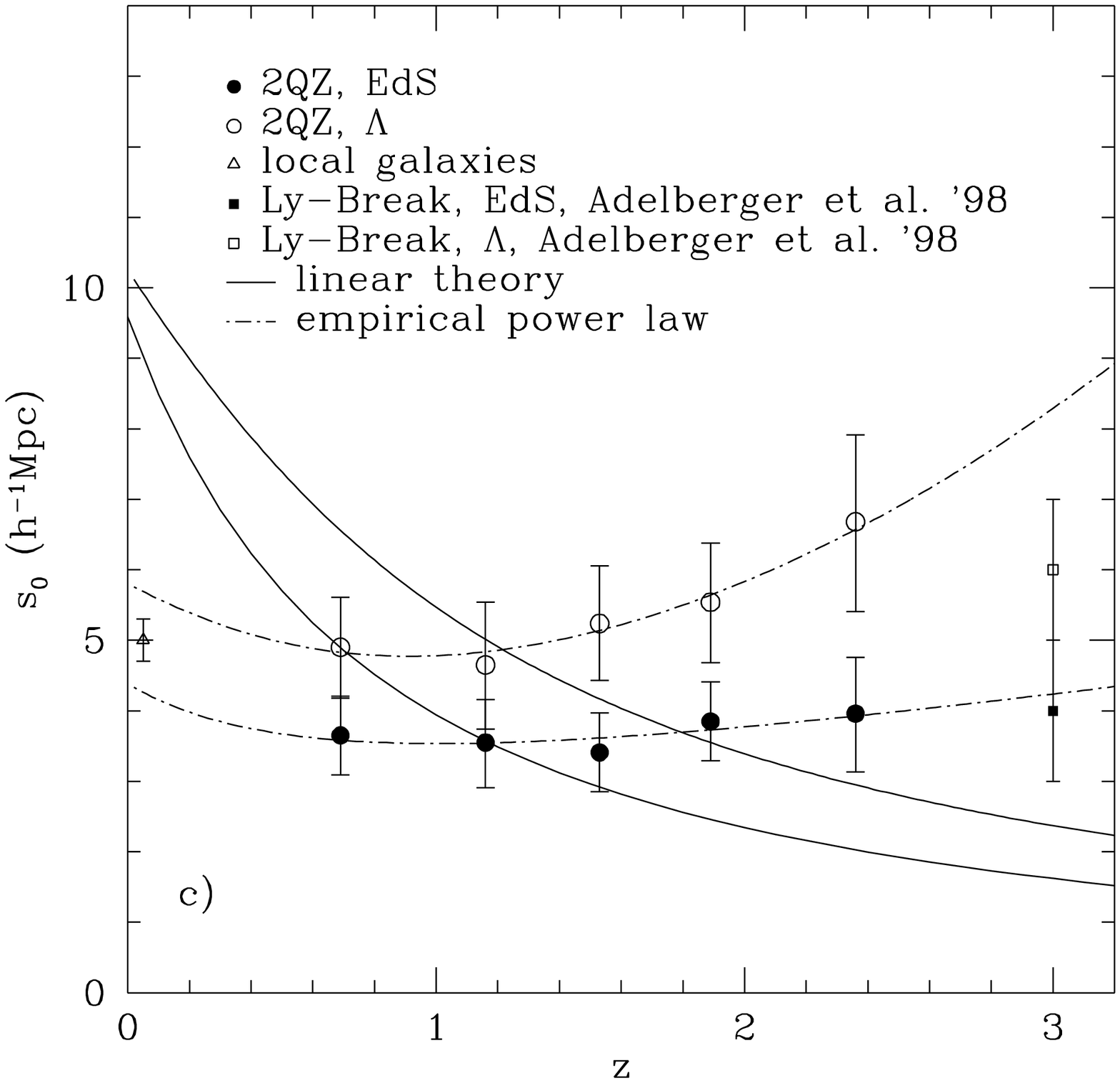}{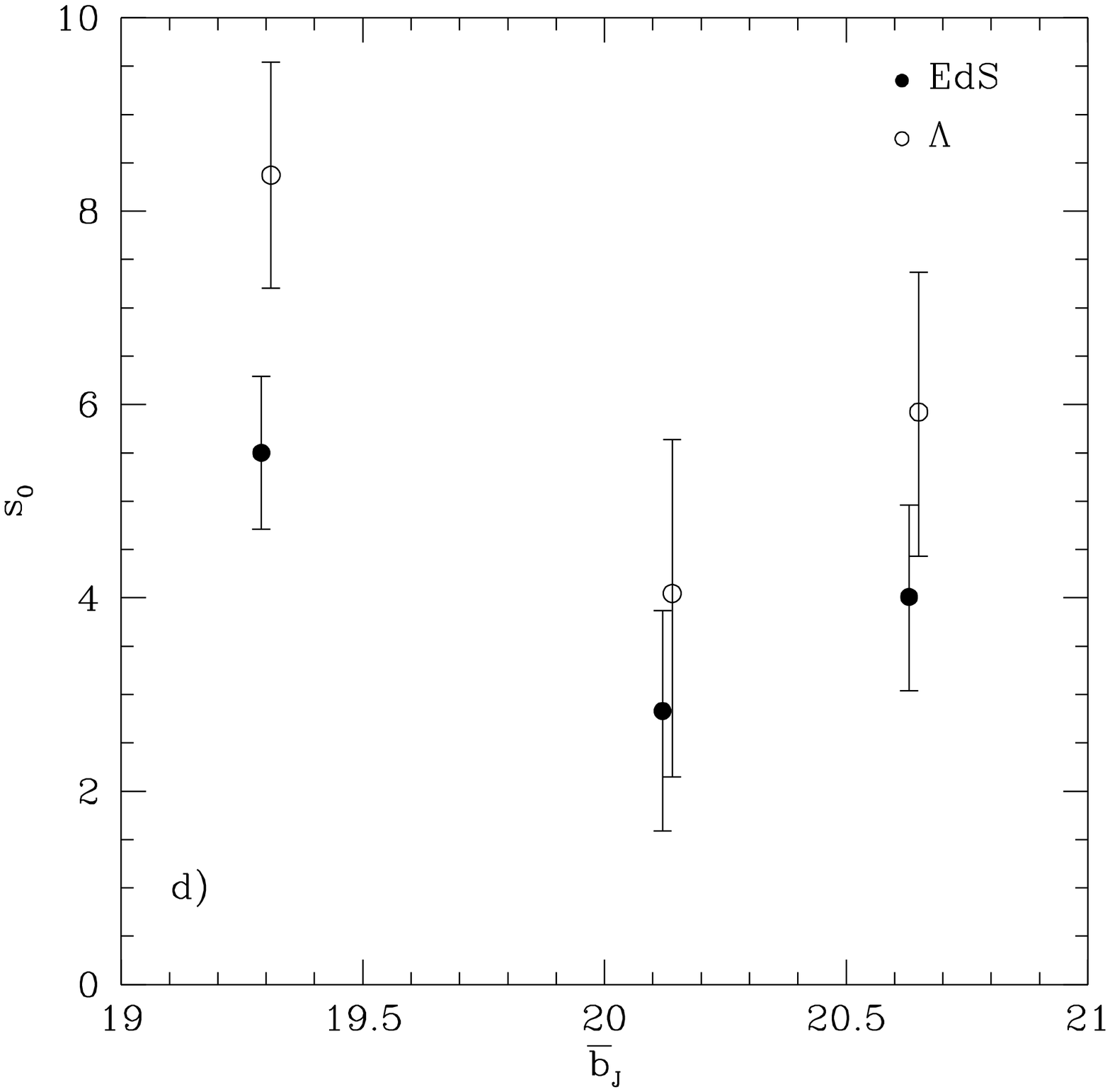}
\caption{QSO clustering results.  a) The two-point correlation
function for 2QZ QSOs in the redshift interval $0.3<z\leq2.9$ in EdS
and $\Lambda$ cosmologies with the best fit power law in each case. b)
A comparison with the  clustering of local galaxies [16,19].
% citation added in by hand as causing a latex error.
The 2QZ data points for the $\Lambda$  model are
omitted for clarity. c) The values of $s_0$ as a function of $z$ for
EdS and $\Lambda$ cosmologies.  We show the linear theory predictions
(solid lines; top: $\Lambda$) and the best fit empirical models
(dot-dashed line).  d) Clustering strength as a function of mean apparent
magnitude.}
\label{fig:xi}
\end{figure}

We have measured the redshift space correlation function,
$\xi\qso(s)$, of 2QZ QSOs, both averaged over the entire sample,
and sub-divided into redshift or magnitude intervals. $\xi\qso(s)$ 
has been estimated assuming two representative cosmological models;
the $\Om=1$ Einstein-de Sitter model (EdS) and a model with $\Om=0.3$
and $\lo=0.7$ ($\Lambda$).  We use the minimum variance \cite{ls93}
correlation function estimator, and details of our method can be found
in Croom et al. (2001) \cite{2qzpaper2}.	

In Fig. \ref{fig:xi}a shows the QSO correlation function for EdS and
$\Lambda$ universes averaged over $0.3<z\leq2.9$, based on the QSOs in
the 10k catalogue.  The amplitude and
shape of $\xi\qso(s)$ is comparable to that of local galaxies
(Fig. \ref{fig:xi}b).  Fitting a standard power law of the form
$\xi\qso(s)=(s/s_0)^{-\gamma}$  we find that
$s_0=3.99^{+0.28}_{-0.34}\Mpc$ and $\gamma=1.55^{+0.10}_{-0.09}$ for
an EdS cosmology.  The effect of a significant cosmological constant
is to increase the separation of QSOs, so that in the $\Lambda$
model the best fit is $s_0=5.69^{+0.42}_{-0.50}\Mpc$ and
$\gamma=1.56^{+0.10}_{-0.09}$.  Both these best fit lines are shown in
Fig \ref{fig:xi}a.

We then sub-divide the 2QZ into five redshift intervals containing
approximately equal number of QSOs.  The correlation function is
measured in each redshift interval separately, and a power law is fit to
the result (assuming the same slope as found from the full
sample).  The resulting clustering scale lengths are shown in Fig. 
\ref{fig:xi}c.  The clustering of QSOs is constant as a function of
redshift over the entire range probed by the 2QZ (EdS).  In the
$\Lambda$ case there is a marginal increase of clustering with
increasing redshift, but a constant value cannot be ruled out.  Making
comparisions to the clustering of high redshift Ly-break galaxies
\cite{adelberger98}, at $z\sim3$, we see that these also have a similar
clustering strength to that of the 2QZ QSOs.  The solid lines in
Fig. \ref{fig:xi}c denote the linear theory evolution of
clustering.  The data disagrees with the linear theory
prediction, implying that QSOs must have a redshift dependent bias
factor, $b\qso(z)$.  We derive an empirical fit to the bias of the
QSOs assuming $b(z)=1+(b(0)-1)G(\Om,\lo,z)^\beta$ where $G(\Om,\lo,z)$
is the linear growth factor (dot-dashed lines in Fig. \ref{fig:xi}c).
The best fit values are $b(0)=1.45^{+0.21}_{-0.16}$ and
$\beta=1.68^{+0.44}_{-0.40}$ (EdS) and $b(0)=1.28^{+0.16}_{-0.11}$ and
$\beta=1.89^{+0.49}_{-0.46}$ ($\Lambda$).  A more detailed analysis is
carried out by Croom et al. (2001) \cite{2qzpaper2}.

We lastly sub-divide our sample into 3 apparent magnitude slices with
equal number of QSOs in each.  The measured values of $s_0$ as a
function of mean apparent magnitude are shown in Fig. \ref{fig:xi}d.
There is no significant difference between the different magnitude
slices, although the brightest QSOs do have marginally stronger
clustering.  We note that sub-dividing on the basis of apparent
magnitude is approximately equvalent to selection relative to $M^*$
due to the extreme luminosity evolution of QSOs over the redshift
range of our sample \cite{2qzpaper1}. 

\section{Cosmological parameters from redshift-space
distortions}\label{sec:zspace}

\begin{figure}
\plottwo{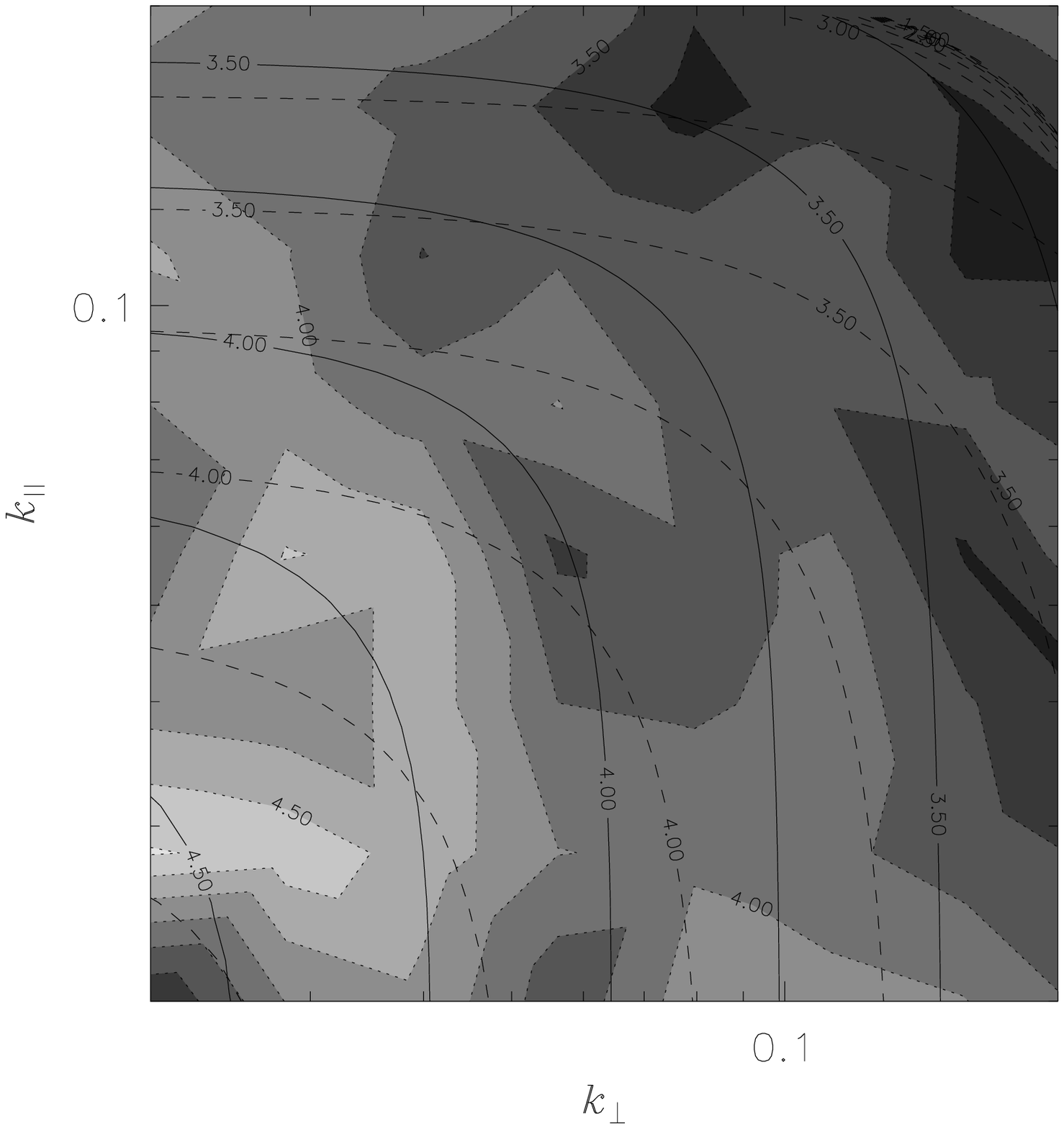}{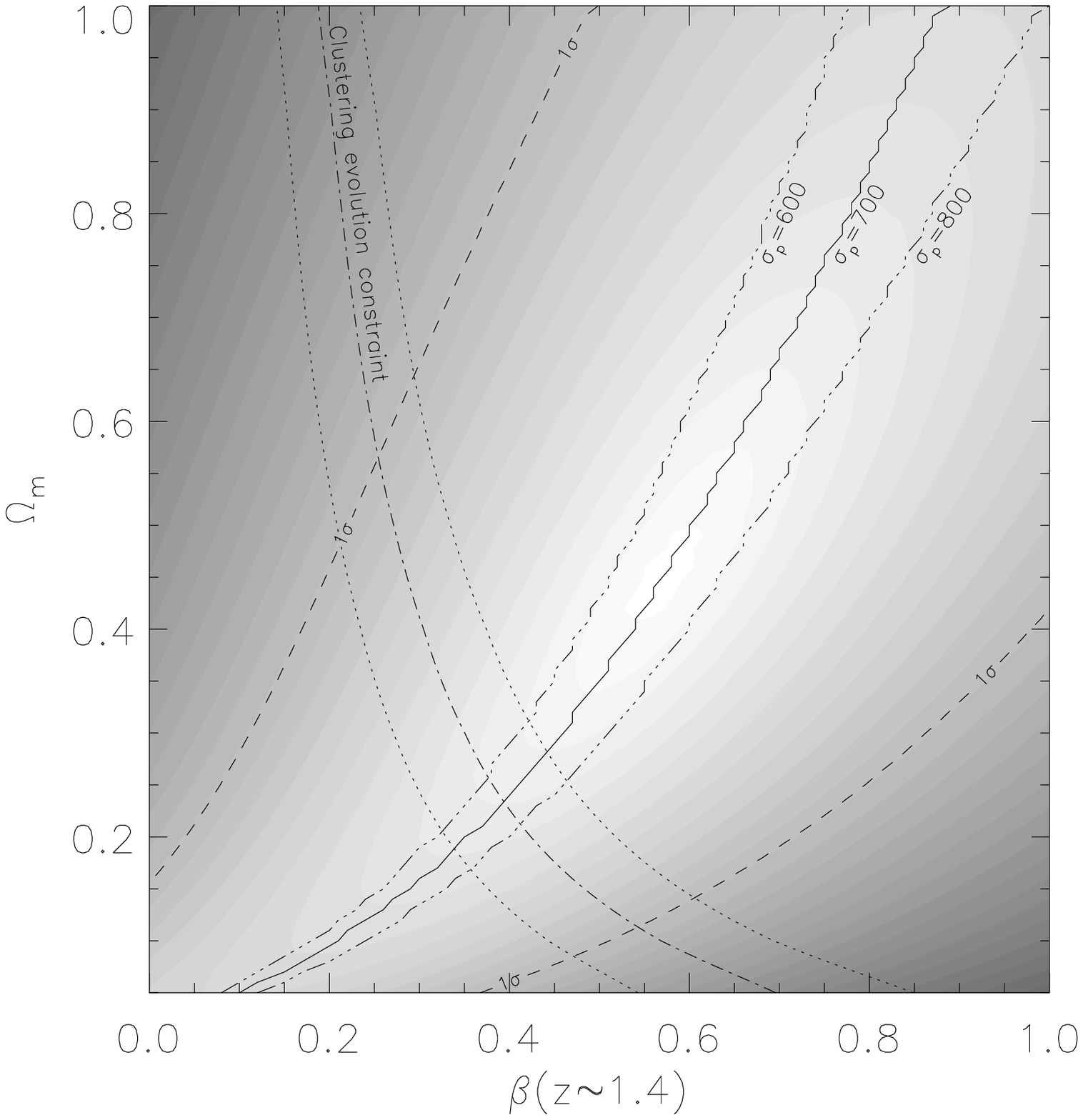}
\caption{Left: $P(k_\parallel,k_\perp)$ estimated from the 2QZ 10k
catalogue. Filled contours of constant $\log(P(k)/\invMpcc)$ are shown
as a function of $k_\parallel/\invMpc$ and $k_\perp/\invMpc$. Overlaid
are the best fit model (solid  contours) with $\beta = 0.39$ and $\Omm
=  0.23$, and a model with $\Omm = 1.0$ and $\beta = 0.19$ (dashed
contours).  Right: filled contours of increasing $\chi^2$ in the
$\Omm$ -- $\beta$ plane for fits to the 2QZ 10k catalogue. The solid
line represents the best fit value of $\beta$ for each $\Omm$ and the
1-$\sigma$ statistical error is given by the dashed line.  The
dot-dot-dot-dash lines above and below the solid line show the  best
fit value of $\beta$ for each $\Omm$ with $\sigp=600\kms$ and
$\sigp=800\kms$ respectively.  Overlaid is the best-fit (dot-dash)
and 1-$\sigma$ (dot) values of  $\beta$ determined using the mass
clustering evolution method.  The joint best fit values obtained are
$\beta = 0.39$ and $\Omm = 0.23$.}
\label{fig:zspace}
\end{figure}

By comparing clustering along and across the line of sight, and
modelling the effects of peculiar velocities (both linear and
non-linear) it is possible to detect the geometric distortions present
if the wrong cosmological model is assumed when determining the clustering
\cite{ap79,bph96}.  At high redshift, the geometric distortion is
particularly sensitive to the cosmological constant $\lo$.  In
practice, a model which also takes into account linear infall is fit
to provide a constraint in the $\lo$ vs. $\beta=\Om^{0.6}/b$ plane.

The QSO power spectrum from the 2QZ 10k sample, measured along and
across the line of sight, $P(k_\parallel,k_\perp)$, is shown in
Fig. \ref{fig:zspace}.  Unfortunately the current 2QZ data only
provides a joint constraint on $\Omega_{\rm m}=1-\lo$ and $\beta$
(shading in Fig. \ref{fig:zspace}b).  Infact there is also some
dependence on the value of the small scale non-linear pair-wise
velocity dispersion, $\sigp$.  Because we are looking at
$P(k_\parallel,k_\perp)$ on large scales this is not a
major issue, but the effects of varying the assumed velocity
dispersion is shown in Fig. \ref{fig:zspace}b.

However, we can obtain a useful constraint on the cosmological world
model by combining the above approach with our knowledge of the linear
growth of clustering.  By using the value of $\beta$ and the
two-point correlation function for nearby galaxies derived from the
2dfGRS \cite{peacock01} we can determine the clustering of matter at
$z=0$ and for a given cosmology we can then derive the clustering of
mass as a function of $z$.  Comparison to the amplitude of QSO
clustering then gives the mean value of the QSO bias and hence
$\beta$.  This approach gives a differing relation between
$\Omega_{\rm m}$ and $\beta$ (dotted and dot-dashed lines in
Fig. \ref{fig:zspace}b), and thus breaks the degeneracy.  The
current best fit derived from this method is
$\Omm=0.23^{+0.44}_{-0.13}$ and $\beta=0.39^{+0.18}_{-0.17}$.
Ths analysis is discussed further in Outram et al., (2001)
\cite{2qzpaper6}.

With the completion of the 2QZ at the end of 2001, analysis of the
complete data set will produce further constraints on cosmological
world models.  This will include different types of analysis than those
discussed here, including the gravitational lensing properties of the
2QZ QSOs.  It is hoped that the 2QZ will become a major resource
for the astronomical community, particularly  when the final data
becomes public at the end of 2002.

\acknowledgements{We warmly thank all the present and former staff of the
Anglo-Australian Observatory for their work in building and operating
the 2dF facility.  The 2QZ is based on observations made with the
Anglo-Australian Telescope and the UK Schmidt Telescope.  NSL is
supported by a PPARC Studentship.  We also thank Marie and Laurence
for organising a very enjoyable and interesting meeting.}

\vfill

\begin{thebibliography}{}{
\bibitem{adelberger98} Adelberger K.~L., Steidel C.~C., Giavalisco M.,
Dickinson M., Pettini M., Kellogg M., 1998, ApJ, 505, 18
\bibitem{ap79} Alcock C., Paczy\'nski B., 1979, Nature, 281, 358
\bibitem{2dfman} Bailey J., Glazebrook K., 1999, 2dF User Manual,
Anglo-Australian Observatory 
\bibitem{bph96} Ballinger W.~E., Peacock J.~A., Heavens A.~F., 1996,
\mnras, 282, 877
\bibitem{bsfp90} Boyle B.~J., Fong R., Shanks T., Peterson B.~A.,
1990, \mnras\, 243, 1 
\bibitem{2qzpaper1} Boyle B.~J.,  Shanks T.,  Croom S.~M.,  Smith
R.~J.,  Miller L., Loaring N., Heymans C.,  2000, \mnras, 317, 1014 
\bibitem{brother99} Brotherton M.~S. et al., 1999, ApJ, 520, L87 
\bibitem{2dfgrs} Colless M., 1999, in Morganti R., Couch W.~J., eds.,
 Proc. ESO/Australia Workshop, Looking Deep in the Southern
 Sky. Springer-Verlag, p.9 
\bibitem{croom} Croom S.~M., 1997, Ph.D. Thesis, University of Durham
\bibitem{2qzpaper2} Croom S.~M.,  Shanks T.,  Boyle B.~J.,  Smith
R.~J.,  Miller L.,  Loaring N., Hoyle F.,  2001, \mnras, 325, 483 
\bibitem{2qzpaper5} Croom S.~M.,  Smith R.~J.,  Boyle B.~J.,  Shanks
T.,  Loaring N.~S.,  Miller L.,    Lewis I.~J.,  2001, \mnras, 322,
L29
\bibitem{ls93} Landy, S.~D., Szalay, A.~S., 1993, ApJ, 412, 64
\bibitem{2dfpaper} Lewis I.~J. et al., 2001, \mnras\ submitted 
\bibitem{2qzpaper6} Outram P.~J., Hoyle F., Shanks T., Boyle B.~J.,
Croom S.~M., Loaring N.~S., Miller L., Smith R.~J., 2001, \mnras, in
press, (astro-ph/0106012) 
\bibitem{peacock01} Peacock, J.~A. et al., 2001, Nature, 410, 169
\bibitem{rspf98} Ratcliffe A., Shanks T., Parker Q.~A., Fong R., 1998,
MNRAS, 296, 173 
\bibitem{smith} Smith R.~J., 1998, Ph.D. Thesis, University of
Cambridge
\bibitem{2qzpaper3} Smith R.~J., Croom S.~M., Boyle B.~J., Shanks T.,
Miller L., Loaring N.~S., 2001, \mnras, submitted
\bibitem{tucker97} Tucker D.~L. et al., 1997, \mnras, 285, L5

}
\end{thebibliography}
\end{document}